\documentclass[aps,pre,preprint,groupedaddress]{revtex4-1}
\usepackage[utf8]{inputenc}
\usepackage[T1]{fontenc}
\usepackage{graphicx}
\usepackage{url}
\usepackage{hyperref}    % przypisy sa linkami
\usepackage{amsmath}

\usepackage{listings}
\begin{document}

\title{Halin graphs are 3-vertex-colorable except even wheels}

\author{A. Kapanowski}

%\email[e-mail: ]{andrzej.kapanowski@uj.edu.pl}
\email[Corresponding author: ]{andrzej.kapanowski@uj.edu.pl}
%\homepage[Visit: ]{Your web page}
%\thanks{}    % przypisy

\author{A. Krawczyk}

\affiliation{Institute of Physics, Jagiellonian University, ulica Łojasiewicza 11, 30-348 Kraków, Poland}

\date{\today}

% Przy revtex 4.1 po abstract jest dopiero maketitle.
% Mozna tez uzywac structured abstracts z punktami.
\begin{abstract}
A Halin graph is a graph obtained by embedding a tree having no nodes
of degree two in the plane, and then adding a cycle to join the leaves
of the tree in such a way that the resulting graph is planar.
According to the four color theorem, Halin graphs are
4-vertex-colorable. On the other hand, they are not 2-vertex-colorable
because they have triangles.
We show that all Halin graphs are 3-vertex-colorable except even wheels.
We also show how to find the perfect elimination ordering
of a chordal completion for a given Halin graph.
The algorithms are implemented in Python using the graphtheory package.
Generators of random Halin graphs (general or cubic) are included.
The source code is available from the public GitHub repository.
\end{abstract}

%\pacs{}               % need showpacs

  % need showkeys

\maketitle

\section{Introduction}
\label{sec:intro}

A Halin graph $H = T \cup C$ is a simple graph obtained by embedding a tree $T$,
which has no nodes of degree two, in the plane. Then new edges connecting
leaves and forming a cycle $C$ are added in such a way that the resulting
graph $H$ is planar.
These graphs were studied by Halin
\cite{1971_Halin}
and Kirkman (cubic graphs)
\cite{1856_Kirkman}.
Every Halin graph is Hamiltonian, 3-connected, and polyhedral.
The simplest Halin graphs with $n$ vertices and $m$ edges are wheels $W_n$,
then $T$ is a star and $C$ is formed from $n-1$ vertices.
The smallest Halin graphs ($n<7$) are $W_4=K_4$, $W_5$, $W_6$, and a 3-prism.
A fan in $H$ is composed of a non-leaf vertex $v$ of $T$, which is adjacent 
to exactly one other non-leaf vertex of $T$, and the leaves adjacent to $v$.
If $H$ is not a wheel then it contains at least two fans
\cite{1983_Cornuejols}.

Many computational problems that are hard on general graphs can be solved
in polynomial time on Halin graphs.
This is attributed to low treewidth equal to three 
\cite{1988_Bodlaender_planar}
and tree-like structure
\cite{1990_Proskurowski_Syslo}.
Contracting a fan to a new single vertex leads to a smaller Halin graph.
Successive fan contractions reduce any Halin graph to a wheel.

Some graph problems on Halin graphs can be solved using dynamic programming
\cite{1988_Bodlaender_dynamic},
in some cases direct algorithms are known.
Our aim is to present direct algorithms for vertex coloring and
for finding a chordal completion of a given Halin graph.
It is shown that all Halin graphs except even wheels can be optimally
colored with three colors.

In this paper unweighted graphs are studied but
let us mention two problems defined on weighted Halin graphs
where the shrinking fans method is used to quickly find a solution.
In 1983 Cornuejols et al. solved the travelling salesman problem
\cite{1983_Cornuejols}.
In 2007 a linear time algorithm finding the maximum-weighted matching 
was shown
\cite{2007_Lu_Li_Lou}.

%There are also parallel algorithms for many problems, such as

The paper is organized as follows. 
Section \ref{sec:recognition} deals with recognition algorithms
for Halin graphs.
In Section \ref{sec:nodecolor}
vertex coloring of Halin graphs is described.
Finding a chordal completion for Halin graphs 
is discussed in Section \ref{sec:chordal_completion}.
In Section \ref{sec:python_implementation} exemplary calculations
for several graph problems are shown.
Conclusions are contained in Section \ref{sec:conclusions}.

\section{Recognition of Halin graphs}
\label{sec:recognition}

Halin graphs can be recognized in linear time.
One can find a planar embedding of a given graph and then choose 
the proper outer face with $m-n+1$ edges
\cite{1983_Syslo_Proskurowski}.
Fomin and Thilikos
\cite{2006_Fomin_Thilikos}
showed that every planar drawing of a Halin graph has from one to four
faces with $n/2+1$ vertices.
It can be also used to recognize Halin graphs.
However, both methods refer to planarity testing which is hard to implement.

In 2016 Eppstein presented a reduction algorithm avoiding planar
embedding of a given graph
\cite{2016_Eppstein}.
Applying two safe reduction rules gradually shrinks the input graph
to the wheel graph, provided the input graph is a Halin graph.
The algorithm was implemented in Python by Eppstein and it was
adapted for the \emph{graphtheory} package
\cite{graphs-dict}.
The algorithm outputs the set of vertices from the outer cycle
and this set is used in next algorithms as a part of the input.
If the set of outer cycle vertices is known then by means of a modified
breadth-first search algorithm (BFS) it is easy to find the inner tree 
for a given Halin graph in linear time.
It is useful to keep the inner tree as a dictionary with parent nodes
and the outer cycle as a doubly linked cyclic list.

\section{Vertex coloring}
\label{sec:nodecolor}

Vertex coloring is a labeling of vertices with colors such that 
no two vertices sharing the same edge have the same color.
The problem of finding the smallest number of colors needed to color
a graph (the chromatic number) is NP-hard in general.
However, in the case of Halin graphs an optimal coloring can be found
in linear time.

A short description of the coloring algorithm is the following.
Let $H = T \cup C$ be a Halin graph.
First, color the tree $T$ with two colors $c_1$ and $c_2$.
Conflicts of colors occur on the cycle $C$.
Second, recolor some vertices from $H$ to resolve conflicts.

\emph{Case 1.} If $C$ is even then recolor every second vertex of $C$
with the third color $c_3$. A color sequence is $[(c_1|c_2)c_3]$,
where $(\cdot|\cdot)$ means an alternative and $[\cdot]$ a repetition.

\emph{Case 2.} If $H$ is an even wheel then recolor $C$ using two colors
$c_3$ and $c_4$. A color sequence is $[c_2c_3]c_4$.
Note that $c_1$ is for the hub.

\emph{Case 3.} If $C$ is odd and cycle nodes have two colors $c_1$ and $c_2$,
then find a color sequence $c_1c_2$. 
Recolor $C$ to get a color sequence $c_1 c_2 c_3 [(c_1|c_2)c_3]$.

\begin{figure}
\centering
\includegraphics[scale=1]{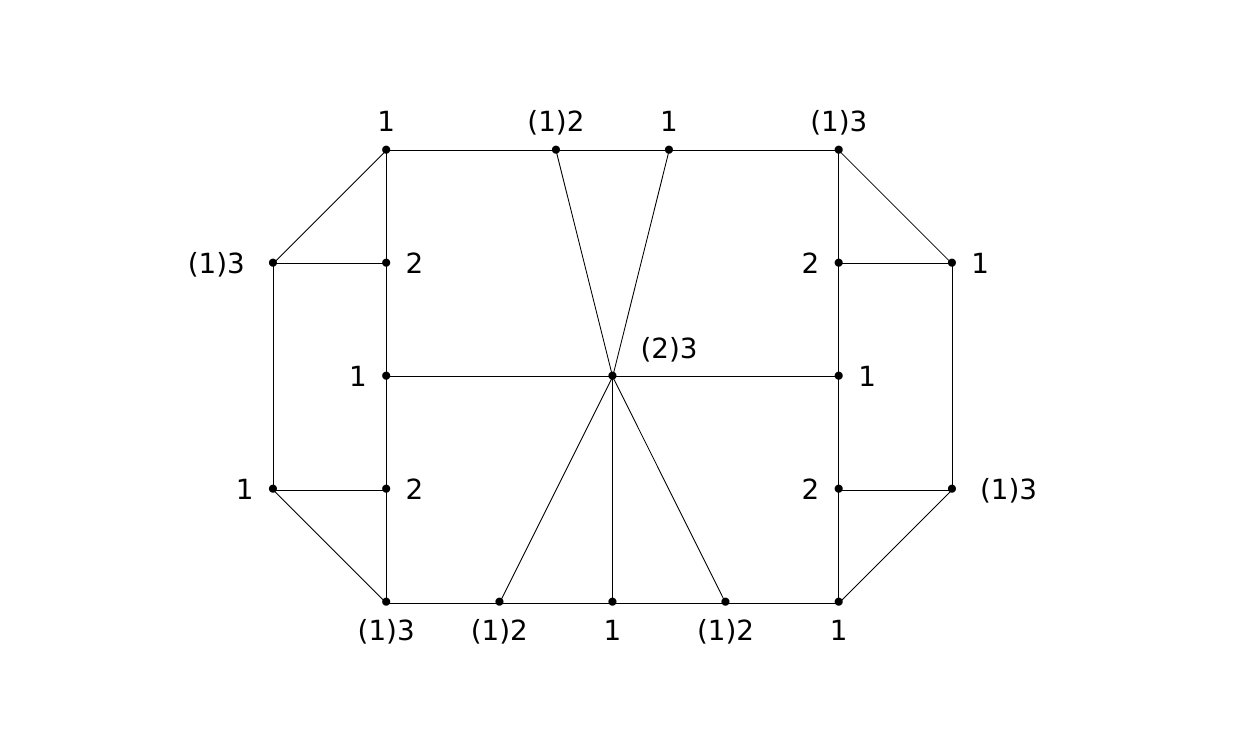}
\caption[Vertex coloring for Halin graphs.]{
\label{fig1}
Vertex coloring for a Halin graph $H = T \cup C$ 
with an odd cycle $C$ (case 4).
After tree $T$ coloring the cycle $C$ has color 1.
An odd pseudo-fan is used during recoloring because all four fans are even.
Old colors are shown in parentheses.
Note that color 2 is present in $C$ in two distant regions.}
\end{figure}

\emph{Case 4.} If $C$ is odd and cycle nodes have one color, say $c_1$.
Then there is an odd fan or pseudo-fan in $H$ with the center color $c_2$.
Recolor $C$ to get a color sequence $c_2 [c_1 c_2] [c_1(c_3|c_2)]c_1 c_3$
and recolor the (pseudo-)fan center with $c_3$.
The sequence  $c_2 [c_1 c_2]$ is for the odd (pseudo-)fan.
The sequence $[c_1(c_3|c_2)]c_1 c_3$ means that 
if the parent node has color $c_2$ then a cycle node is recolored with $c_3$ but 
if the parent node has color $c_3$ then a cycle node is recolored with $c_2$
(see Figure \ref{fig1}).

The tree $T$ coloring runs in $O(n)$ time because this is BFS.
The cycle $C$ coloring runs in time proportional to the cycle length
and finding an odd (pseudo-)fan (case 4) is done in the same time.
Therefore, the overall running time is $O(n)$.

\section{Chordal completion}
\label{sec:chordal_completion}

A chordal completion of a Halin graph $H$ is a chordal graph $G$
on the same vertex set $V(G) = V(H)$ such that $H$ is a subgraph of $G$.
We are interested in a chordal completion with the treewidth equal to three
because this is the treewidth of all Halin graphs
\cite{1988_Bodlaender_planar}.
Every chordal graph has a perfect elimination ordering (PEO).
This is a vertex ordering such that, for each vertex $v$,
$v$ with his neighbors that occur after $v$ in the order form a clique.
We will show how to find a PEO for a chordal completion of a given
Halin graph.

\begin{figure}
\centering
\includegraphics[scale=1]{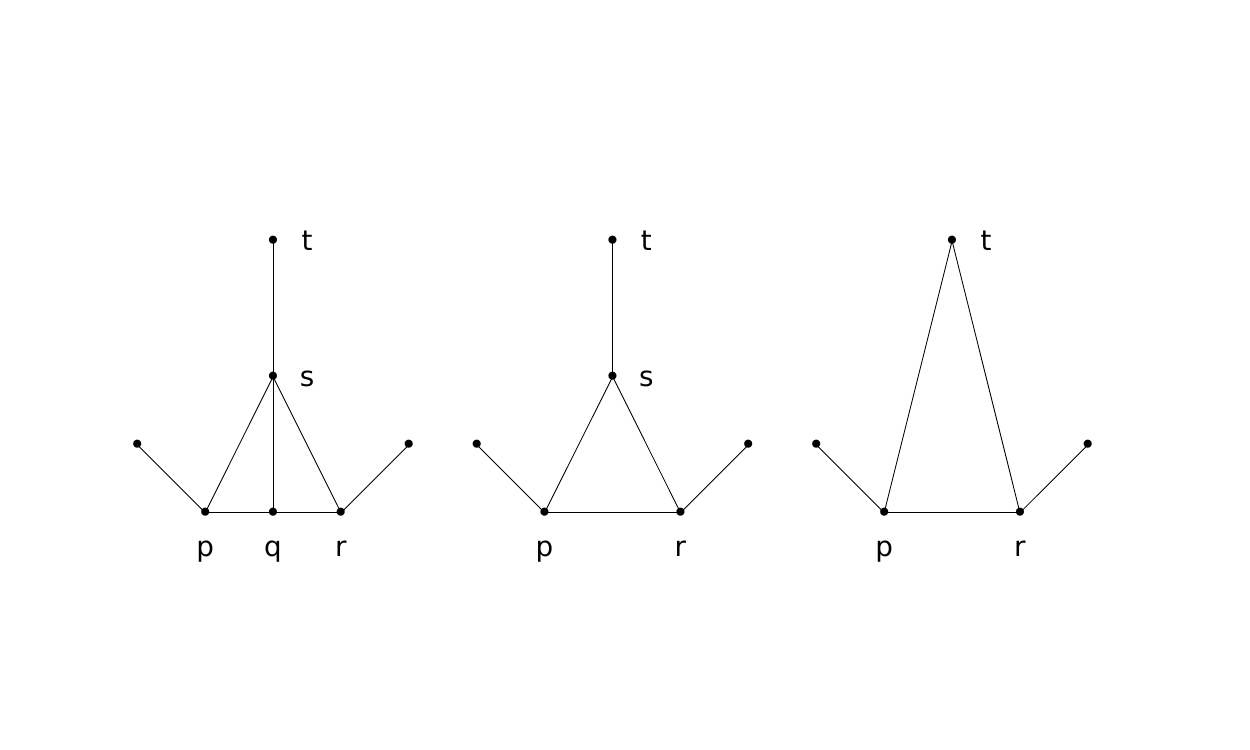}
\caption[Finding PEO for a chordal completion.]{
\label{fig2}
Finding PEO for a chordal completion of a Halin graph $H = T \cup C$.
Left: a small fan with three vertices $p$, $q$, and $r$ in $C$.
Middle: the fan after a reduction \textbf{R1}, $q$ is removed,
a new edge $pr$ is added.
Right: the fan after a reduction \textbf{R2}, $s$ is removed,
new edges $pt$ and $rt$ are added.}
\end{figure}

The algorithm takes a Halin graph $H$ and the set of his outer nodes as
an input. The following two reduction rules are used in order to reduce
$H$ to the four-vertex complete graph $K_4$
(see Figure \ref{fig2}).

\textbf{R1.} Let $p$, $q$, and $r$ be three vertices that induce a path,
and this a part of a fan or a wheel with the center $s$.
Add a new edge $pr$ to form a clique $\{p,q,r,s \}$.
Delete $q$ from the graph and add $q$ to PEO.

\textbf{R2.} Let $p$ and $r$ be two vertices that form a small fan
with the center in $s$. The third neighbor of $s$ is $t$.
Add two new edges $pt$ and $rt$ to form a clique $\{ p,r,s,t \}$.
Delete $s$ from the graph and add $s$ to PEO.

The reduction rules lead to $K_4$ and the vertices from $K_4$
should be added to PEO in any order.
Note that after \textbf{R1} or \textbf{R2} a smaller Halin graph
is obtained.
There are many possible reduction sequences and it corresponds to 
many possible PEO of different chordal completions.
New edges added during reduction steps can be recorded and used to build
the complete chordal supergraph of H.

Removing a node and adding it to PEO runs in constant time.
The only problem is in finding fans until a graph will become a wheel.
The algorithm runs along the cycle, finds both ends of a node sequence,
removes the nodes between the ends, removes the center of the fan,
and then repeats these steps. For large graphs the algorithm wraps
several times around the inner tree.
Our theoretical estimates for necklace graphs lead to $O(n^2)$ running time
but computer experiments on random Halin graphs suggest overall 
linear-time complexity.

\section{Python implementation}
\label{sec:python_implementation}

The algorithms presented in this work are implemented 
in the Python programming language
\cite{python}
using the \emph{graphtheory} package. It can be installed
from the GitHub repository using Git, a distributed version control system
\cite{2016_Kapanowski}.
The package uses consistent graph interface, provides readable source code,
and the real computational complexity is in agreement with the theory.
The package has a growing number of available algorithms
and graph generators, wheels and necklace graphs for example.
Let us show exemplary calculations for Halin graphs.

\begin{lstlisting}
>>> from graphtheory.structures.edges import Edge
>>> from graphtheory.structures.graphs import Graph
>>> from graphtheory.planarity.halintools import *
>>> from graphtheory.planarity.halin import HalinGraph
>>> from graphtheory.planarity.halinnodecolor import HalinNodeColoring
>>> from graphtheory.planarity.halinpeo import HalinGraphPEO
>>> G = make_halin(10)   # random Halin graph with 10 vertices
# G = make_halin_cubic(10)   # random cubic Halin graph
# Recognition.
>>> algorithm = HalinGraph(G)
>>> algorithm.run()
>>> outer = algorithm.outer   # the outer cycle found
# Vertex coloring.
>>> algorithm = HalinNodeColoring(G, outer)
>>> algorithm.run()
>>> print algorithm.parent   # the inner tree as a dict
>>> print algorithm.color   # a dict with node colors
# Chordal completion.
>>> algorithm = HalinGraphPEO(G, outer)
>>> algorithm.run()
>>> print algorithm.parent   # the inner tree as a dict
>>> print algorithm.order   # PEO
\end{lstlisting}

\section{Conclusions}
\label{sec:conclusions}

Halin graphs are very interesting from the algorithmic point of view
because they fall between trees and general planar graphs.
Some graph problems remain hard even when their domains are restricted
to planar graphs (3-colorability, node cover, independent set).
On the other hand, for trees polynomial-time algorithms are known
for almost all problems.
That is why many studies were devoted to Halin graphs and many
polynomial-time or even linear-time algorithms were described.

In this paper we showed that all Halin graphs are 3-vertex-colorable 
except even wheels. Our algorithm follow the definition of Halin graphs
and first it colors an inner tree, then it recolors some vertices
to remove conflicts. 
A different approach was described in a paper by Kavin at al.
\cite{2014_Kavin}.

The second algorithm presented in this paper finds a perfect elimination
ordering of a chordal completion for a given Halin graph.
It is also possible to find the complete chordal subgraph
with treewidth equal to three. Two reduction rules are introduced
which make use of a recursive structure based on fans and wheels.

%\listoffigures

%\newpage

%\listoftables

%\newpage

\end{document}